\documentclass[12pt]{article}

\newcommand{\be}{\begin{equation}}
\newcommand{\ee}{\end{equation}}
\newcommand{\bi}[1]{\vspace{-3mm} \bibitem{#1}}

\begin{document}

\begin{center}
Physics Letters A. Vol.341. N.5/6. (2005) pp.467-472.
\end{center}

\begin{center}
{\Large \bf Possible Experimental Test of \\
Continuous Medium Model for Fractal Media}
\vskip 5 mm

{\large \bf Vasily E. Tarasov } \\

\vskip 3mm
{\it Skobeltsyn Institute of Nuclear Physics, \\
Moscow State University, Moscow 119992, Russia}

{E-mail: tarasov@theory.sinp.msu.ru}
\end{center}
\vskip 11 mm

\begin{abstract}
We use the fractional integrals to describe fractal media.
We consider the fractal media as special 
("fractional") continuous media.  
We discuss the possible experimental testing of the continuous medium
model for fractal media that is suggested in 
Phys. Lett. A. 336 (2005) 167-174.
This test is connected with measure of period of 
the Maxwell pendulum with fractal medium cylinder.

\end{abstract}



\section{Introduction}

The real structure of the media are characterized 
by an extremely complex and irregular geometry \cite{Mand}. 
Because the methods of Euclidean geometry, which ordinarily 
deals with regular sets, are purely suited for
describing objects such as in nature, stochastic
models are taken into account \cite{Beran}. 
Another possible way of describing a complex structure of 
the media is to use fractal theory of sets
of fractional dimensionality \cite{Mand}.
Although the fractal dimensionality does
not reflect completely the geometric 
properties of the fractal, it nevertheless permits
a number of important conclusions about the behavior 
of fractal structures \cite{Zaslavsky1,Zaslavsky2}. 
For example, if it is
assumed that matter with a constant density is distributed 
over the fractal, then the mass of the fractal enclosed 
in a volume of characteristic dimension
$R$ satisfies the scaling law $ M (R) \sim R^{D}$, whereas
for a regular n-dimensional Euclidean object $M (R) \sim R^n$.
Let us assume that a medium can be treated on a scale $R$ as 
a stochastic fractal of dimensionality $D <3$ embedded in a Euclidean
space of dimensionality $n = 3$. Naturally, in real objects 
the fractal structure cannot be observed on all
scales.  
For example, Katz and Thompson \cite{KT} presented experimental 
evidence indicating that the pore spaces of a set of
sandstone samples are fractals 
in length extending from 10 angstrom to 100 $\mu m$. 


In the general case, the fractal media cannot 
be considered as continuous media.
There are points and domains that are not filled of the medium particles.
We suggest to consider the fractal media  as special 
(fractional) continuous media.
We use the procedure of replacement of the fractal medium 
with fractal mass dimension by some continuous medium that 
is described by fractional integrals.
This procedure is a fractional generalization of 
Christensen approach \cite{Chr}.
Suggested procedure leads to the fractional integration and 
differentiation to describe fractal media.
The fractional integrals allow us to take into account the fractality
of the media.
In order to describe the fractal media by continuous medium model, 
we must use the fractional integrals. 
The order of fractional integral is equal 
to the fractal mass dimension of the media.
More consistent approach to describe the fractal media
is connected with the mathematical definition of the integrals
on fractals. In Ref. \cite{RLWQ} was proved that integrals 
on net of fractals can be approximated by fractional integrals. 
Therefore, we can consider the fractional integrals as approximation 
of the integrals on fractals \cite{Nig4}.
In Ref. \cite{chaos}, we proved that fractional integrals 
can be considered as integrals over the space with fractional 
dimension up to numerical factor. To prove this statement, 
we use the well-known formulas of dimensional regularizations \cite{Col}.

We use the fractional integrals to describe fractal media.
We consider the simple example of application of fractal integrals to
describe the fractal medium.
We discuss the possible experimental testing of the continuous medium
model for fractal media that is suggested in \cite{PLA05}.
This test is connected with measure of period of 
the Maxwell pendulum with fractal medium cylinder.

It is known that the homogeneous medium cylinder with mass $M$ and 
the radius $R$ such that the medium of cylinder has the integer mass 
dimension, has the moments of inertia 
\[ I^{(2)}_z=(1/2)MR^2 ,\] 
where $z$ is a cylinder axis.
If the fractal medium cylinder has the mass $M$, radius $R$,
and the fractal mass dimension, then the 
moment of inertia has the form
\be \label{Ialpha} I^{(\alpha)}_z=\frac{\alpha}{\alpha+2} MR^2 , \ee
where $\alpha$ is a fractal mass dimension of cross-section of
the cylinder ($1<\alpha\le 2$). The parameter $\alpha$ can be calculated 
by box counting method for the cross-section of the cylinder.

As the result, we get that two homogeneous cylinders 
with the same masses and radiuses have the unequal moments of inertia. 
These cylinders have the different properties as Maxwell pendulums.
The periods of oscillation for these pendulums are connected by the equation
\[ \Bigl(\frac{T^{(\alpha)}_0}{T^{(2)}_0}\Bigr)^2=
\frac{4(\alpha+1)}{3(\alpha+2)} . \]
where $T^{(\alpha)}_0$ is a period of oscillation of 
fractal medium Maxwell pendulum.
For example, the experiment can be realized by using the sandstone cylinder.
Note that Katz and Thompson \cite{KT} presented experimental 
evidence indicating that the pore spaces of a set of sandstone samples 
are fractals and are self-similar over three to four orders of magnitude 
in length extending from 10 angstrom to 100 $\mu m$.

\section{Fractal Media and Fractional Integrals}

The cornerstone of fractals is the meaning of dimension, 
specifically the fractal dimension. 
Fractal dimension can be best calculated by box counting 
method which means drawing a box of size $R$ 
and counting the mass inside. 
The mass fractal dimension \cite{Mand,Schr} can be easy measured
for fractal media.
The properties of the fractal media like mass obeys a power law relation
\be \label{MR} M \sim R^{D} , \quad (D<3),  \ee
where $M$ is the mass of fractal medium, $R$ is a box size 
(or a sphere radius),
and $D$ is a mass fractal dimension. 
Amount of mass of a medium inside a box of size $R$
has a power law relation (\ref{MR}).

The power law relation (\ref{MR}) can be naturally 
derived by using the fractional integral.
In Ref \cite{PLA05}, we prove that the mass fractal dimension 
is connected with the order of fractional integrals. 
Therefore the fractional integrals can be used to describe fractal
media with non-integer mass dimensions. 

Let us consider the region $W_A$ in 3-dimensional 
Euclidean space $E^3$, where $A$ is the midpoint of this region.
The volume of the region $W_A$ is denoted by $V(W_A)$.
If the region $W_A$ is a ball with the radius $R_A$,
then the midpoint $A$ is a center of the ball, and
the volume $V(W_A)=(4/3)\pi R^3_A$ .
The mass of the region $W_A$ in the fractal medium is denoted by $M(W_A)$. 
The fractality of medium means that the mass of this medium 
in any region $W_A$ of Euclidean space $E^3$ increase more slowly 
that the volume of this region.
For the ball region of the fractal medium, 
this property can be described by the power law (\ref{MR}), 
where $R$ is the radius of the ball $W_A$. 

Fractal media are called homogeneous fractal media if the power 
law (\ref{MR}) does not depend on the translation  
of the region. The homogeneity property of the medium
can be formulated in the form:
For all regions $W_A$ and $W_B$ of the homogeneous fractal medium 
such that the volumes are equal $V(W_A)=V(W_B)$, 
we have that the masses of these regions are equal $M(W_A)=M(W_B)$. 
Note that the wide class of the fractal 
media satisfies the homogeneous property.
In many cases, we can consider the porous media \cite{Por1,Por2}, 
polymers \cite{P}, colloid aggregates \cite{CA}, and 
aerogels \cite{aero} as homogeneous fractal media.

In Ref. \cite{PLA05}, the continuous medium model for the fractal media was
suggested. Note that the fractality and 
homogeneity properties can be realized in the following forms: \\

\noindent
(1) Fractality:
The mass of the ball region $W$ of fractal medium obeys a power law relation
\be \label{MR2} M_D(W)=M_0 \Bigl( \frac{R}{R_c} \Bigr)^D , \ee
where $D<3$, $R$ is the radius of the ball, and
$R_c$ is the characteristic value. 
In the general case, we have the scaling law relation
$M_D(\lambda W)=\lambda^D M_D(W)$, 
where $\lambda W=\{\lambda {\bf x}, \ \ {\bf x} \in W \}$. \\

\noindent
(2) Homogeneity:
The local density of homogeneous fractal media
is translation invariant value that have the form
$\rho({\bf x})=\rho_0=const$.

\vskip 3mm

These requirements can be realized by the 
fractional generalization of the equation
\be \label{MW} M_3(W)=\int_W \rho({\bf x}) d V_3  . \ee
Let us define the fractional integral
in Euclidean space $E^3$ in the Riesz form \cite{SKM}.
The fractional generalization of Eq. (\ref{MW}) can be realized
in the following form
\be \label{ID} M_D(W)=\int_W \rho({\bf x}) dV_D , \ee
where $dV_D=c_3(D,{\bf x})d V_3$, and 
\[ c_3(D,{\bf x})=\frac{2^{3-D} \Gamma(3/2)}{\Gamma(D/2)} |{\bf x}|^{D-3} . \]
Here we use the initial points in the fractional integrals are set to zero.
The numerical factor in Eq. (\ref{ID}) has this form in order to
derive usual integral in the limit $D\rightarrow (3-0)$.
Note that the usual numerical factor
$\gamma^{-1}_3(D)={\Gamma(1/2)}/{2^D \pi^{3/2} \Gamma(D/2)}$,
which is used in Ref. \cite{SKM} 
leads to $\gamma^{-1}_3(3-0)= {\Gamma(1/2)}/{2^3 \pi^{3/2} \Gamma(3/2)}$ 
in the limit $D\rightarrow (3-0)$. 

In order to have the usual dimensions of the physical values,
we can use vector ${\bf x}$, and coordinates 
$x$, $y$, $z$ as dimensionless values.
For example, we can define $x=x_1/x_0$, $y=x_2/x_0$, $z=x_3/x_0$, where 
$x_1$, $x_2$, and $x_3$ are the coordinates with the usual physical dimension.

We can rewrite Eq. (\ref{ID}) in the form
\be \label{MWD}  M_D(W)=\frac{2^{3-D} \Gamma(3/2)}{\Gamma(D/2)}
\int_W \rho({\bf x}) |{\bf x}|^{D-3} d^3 x . \ee

If we consider the homogeneous fractal medium 
($\rho({\bf x})=\rho_0=const$) and the ball region 
$W=\{{\bf x}: \  |{\bf x}|\le R \}$, then we have 
\[ M_D(W)= \rho_0 \frac{2^{3-D} \Gamma(3/2)}{\Gamma(D/2)} 
\int_W |{\bf x}|^{D-3} d V_3 . \]
Using the spherical coordinates, we get
\[ M_D(W)= \frac{\pi 2^{5-D} \Gamma(3/2)}{\Gamma(D/2)} \rho_0 
\int_W |{\bf x}|^{D-1} d |{\bf x}|= 
\frac{2^{5-D} \pi \Gamma(3/2)}{D \Gamma(D/2)} \rho_0 R^{D} . \]
As the result, we have $M(W)\sim R^D$, i.e., we derive Eq. (\ref{MR2})
up to the numerical factor.
Therefore the fractal medium with non-integer mass dimension $D$ can be
described by fractional integral of order $D$.

Note that the interpretation of the fractional integration
is connected with fractional dimension \cite{chaos}.
This interpretation follows from
the well-known formulas for dimensional regularizations \cite{Col}:
The fractional integral can be considered as a 
integral in the fractional dimension space up to the numerical 
factor $\Gamma(D/2) /( 2 \pi^{D/2} \Gamma(D))$.

\section{Moment of Inertia of Fractal Medium Cylinder}

In this section, we consider the fractional generalization of the equation
for moment of inertia. 

The equation for the moment of inertia of homogeneous cylinder with
integer mass dimension has the well-known form
\be \label{0} 
I^{(2)}_z=\rho_0 \int _S (x^2+y^2) dS_{2} \int_L dz . \ee
Here $z$ is the cylinder axis, and $dS_2=dxdy$. 
The fractional generalization of Eq. (\ref{0}) for moment of inertia
can be defined by the equation
\be \label{1} 
I^{(\alpha)}_z=\rho_0 \int _S (x^2+y^2) dS_{\alpha} \int_L dl_{\beta} , \ee
where we use the following notations
\be \label{2} dS_{\alpha}=c(\alpha) 
\left(\sqrt{x^2+y^2}\right)^{\alpha-2} dS_2 , 
\quad dS_2=dxdy , 
\quad c(\alpha)=\frac{2^{2-\alpha}}{\Gamma(\alpha/2)} , \quad
dl_{\beta}= \frac{|z|^{\beta-1}}{\Gamma(\beta)} dz . \ee
The numerical factor in Eq. (\ref{1}) has this form in order to
derive usual integral in the limit $\alpha\rightarrow (2-0)$ and 
$\beta \rightarrow (1-0)$.
The parameters $\alpha$ and $\beta$ are
\[ 1 < \alpha \le 2, \quad 0 < \beta \le 1 . \]
If $\alpha=2$ and $\beta=1$, then Eq. (\ref{1}) has form (\ref{0}).
The parameter $\alpha$ is a fractal mass dimension of the cross-section 
of cylinder. This parameter can be easy calculated from
the experimental data. It can be calculated by box counting method 
for the cross-section of the cylinder.

Substituting Eq. (\ref{2}) in Eq. (\ref{1}), we get
\[ I^{(\alpha)}_z=
\frac{\rho_0 c(\alpha)}{\Gamma(\beta)} \int_S (x^2+y^2)^{\alpha/2} dS_2 
\int^H_0 z^{\beta-1} dz . \]
Here we consider the cylindrical region $W$ that is defined by the relations
\be \label{cyl}  L=\{z: \ 0\le z\le H \}, 
\quad S=\{(x,y): \ 0\le x^2+y^2 \le R^2 \} . \ee
Using the cylindrical coordinates $(\phi,r,z)$, we have
\[ dS_2=dxdy=r dr d\phi , \quad (x^2+y^2)^{\alpha/2}=r^{\alpha} . \]
Therefore the moment of inertia is defined by
\be \label{11} I^{(\alpha)}_z=\frac{2 \pi \rho_0 c(\alpha)}{\Gamma(\beta)} 
\int^R_0 r^{\alpha+1} dr \int^H_0 z^{\beta-1} dz=  
\frac{2 \pi \rho_0 c(\alpha)}{(\alpha+2) \beta \Gamma(\beta)} 
R^{\alpha+2} H^{\beta} . \ee
As the result, we have 
\[ I^{(\alpha)}_z=
\frac{2 \pi \rho_0 c(\alpha)}{(\alpha+2) 
\beta \Gamma(\beta)} R^{\alpha+2} H^{\beta} . \]
If $\alpha=2$ and $\beta=1$, we get $I^{(2)}_z=(1/2)\pi\rho_0 R^4H$.

The mass of the usual medium cylinder (\ref{cyl}) is defined by the equation
\be \label{M2} M=\rho_0 \int_S dS_{2} \int_L dz= 
2 \pi \rho_0  \int^R_0 r dr \int^H_0 dz=  \pi \rho_0 R^2 H . \ee
We can consider the fractional generalization of this equation.
The mass of fractal medium cylinder (\ref{cyl}) can be defined 
by the equation
\be \label{Ma} M_{\alpha}=\rho_0 \int_S dS_{\alpha} \int_L dl_{\beta} . \ee
Using the cylinder coordinates, we get 
\[ M_{\alpha}=\frac{2 \pi \rho_0 c(\alpha)}{\Gamma(\beta)} \int^R_0 r^{\alpha-1} dr 
\int^H_0 z^{\beta-1} dz=  
\frac{2 \pi \rho_0 c(\alpha)}{\alpha \beta \Gamma(\beta)} R^{\alpha} H^{\beta} . \]
As the result, we have
\be \label{M-F} 
M_{\alpha}=\frac{2 \pi \rho_0 c(\alpha)}{\alpha \beta \Gamma(\beta)} R^{\alpha} H^{\beta} . \ee

Substituting the mass (\ref{M-F}) in the moment of inertia (\ref{11}), 
we get the relation
\be \label{I-F} I^{(\alpha)}_z=\frac{\alpha}{\alpha+2} M_{\alpha} R^2 . \ee
Note that Eq. (\ref{I-F}) does not have the parameter $\beta$. 
If $\alpha=2$, we have the well-known relation $I^{(2)}_z=(1/2) M R^2$ 
for the homogeneous cylinder that has the integer mass dimension $D=3$
and $\alpha=2$.

If we consider the fractal medium cylinder with the mass and radius 
that are equal to mass and radius of 
the homogeneous medium cylinder with integer mass dimension, 
then the moments of inertia of these cylinders are connected by the equation
\be I^{(\alpha)}_z=\frac{2\alpha}{\alpha+2} I^{(2)}_z .\ee
Here $I^{(2)}_z$ is the moment of inertia for the
cylinder with integer mass dimension $D=3$ and $\alpha=2$.
For example, the parameter $\alpha=1.5$ leads us to the relation
$I^{(3/2)}=(6/7) I^{(2)}_z$.
Using $1\le \alpha \le 2$, we have the general relation
\[ (2/3) \le I^{(\alpha)}_z/ I^{(2)}_z \le 1 . \]

\section{Equation of Motion for Maxwell Pendulum}

Maxwell pendulum is used to demonstrate transformations 
between gravitational potential energy and rotational kinetic energy.
Wind the string up on the small axis, giving the device some initial 
gravitational potential energy. When released, this gravitational 
potential energy is converted into rotational kinetic energy, 
with a lesser amount of translational kinetic energy.
We consider the Maxwell pendulum as a cylinder that is suspended
by string. The string is wound on the cylinder.

The equations of motion for Maxwell pendulum have the form
\[ M_{\alpha} \frac{dv_y}{dt}=M_{\alpha}g-T , \quad 
I^{(\alpha)}_z\frac{d\omega_z}{dt}=R T , \]
where $g$ is the acceleration such that $g\simeq 9.81 (m/s^2)$;
the axis $z$ is a cylinder axis, $T$ is a string tension. 
Using $v_y=\omega_z R$, we have
\[ M\frac{dv_y}{dt}=Mg-\frac{I^{(\alpha)}_z}{R^2} \frac{dv_y}{dt} . \] 
As the result, we get the acceleration of the cylinder
\be \label{a-y} a_y=\frac{dv_y}{dt}=\frac{Mg}{M+I^{(\alpha)}_z/R^2}
=\frac{g}{1+I^{(\alpha)}_z/(MR^2)} .  \ee
Substituting Eq. (\ref{I-F}) in Eq. (\ref{a-y}), we get
\be a_y=\frac{\alpha+2}{2\alpha+2} g . \ee
For the fractal mass dimension of cross-section of cylinder
$\alpha=1.5$, we get $a_y=(3/5)g \simeq 6.87 \ (m/s^2)$.
For the cylinder with integer mass dimension of 
the cross-section ($\alpha=2$), 
we have $a_y=(2/3)g \simeq 6.54 \ (m/s^2)$.
The period $T_0$ of oscillation for this Maxwell pendulum 
is defined by the equation
\[ T_0=4t_0=4\sqrt{2L/a_y}, \]
where $L$ is a string length, and 
the time $t_0$ satisfies the equation  $a_yt^2_0/2=L$.
Therefore, we get the relation for the periods
\be \frac{T^{(\alpha)}_0}{T^{(2)}_0}=\sqrt{{a^{(2)}_y}/{a^{(\alpha)}_y}}=
\sqrt{\frac{4(\alpha+1)}{3(\alpha+2)}} . \ee
Using $1< \alpha <2$, we can see that 
\[ (8/9)<\Bigl(T^{(\alpha)}_0 / T^{(2)}_0\Bigr)^2<1 . \]
Note the parameter $\alpha$ can be calculated by box counting method
for the cross-section of the cylinder.
For $\alpha=1.5$, we have $\Bigl({T^{(\alpha)}_0}/{T^{(2)}_0}\Bigr)^2=0.952$.

\newpage

\section{Conclusion}

In the general case, the fractal media cannot 
be considered as continuous media.
There are points and domains that are not filled of the medium particles.
In Ref. \cite{PLA05}, we suggest to consider the fractal media  
as special ("fractional") continuous media.
We use the procedure of replacement of the medium 
with fractal mass dimension by some continuous medium that 
is described by fractional integrals.
This procedure is a fractional generalization of 
Christensen approach \cite{Chr}.
Suggested procedure leads to the fractional integration and 
differentiation to describe fractal media.
The fractional integrals are considered as approximation 
of the integrals on fractals \cite{RLWQ}. Note that 
fractional integrals can be considered as integrals 
over the space with fractional dimension up to numerical factor 
\cite{chaos,PRE05,JPCS}. 
To prove we use the well-known 
formulas of dimensional regularizations \cite{Col}. 
The fractional integrals allow us to take into account 
the fractality of the media.

The fractional continuous models of fractal media
can have a wide application. 
This is due in part to the relatively small numbers of parameters 
that define a random fractal medium of great complexity
and rich structure.
In many cases, the real fractal structure of matter 
can be disregarded and the medium can be replaced by  
some "fractional" continuous mathematical model. 
In order to describe the media with 
non-integer mass dimension, we must use the fractional calculus.
Smoothing of the microscopic characteristics over the 
physically infinitesimal volume transform the initial 
fractal medium into "fractional" continuous model
that uses the fractional integrals. 
The order of fractional integral is equal 
to the fractal mass dimension of the medium.
The fractional continuous model allows us
to describe dynamics for wide class fractal media 
\cite{AP05,Chaos05,Physica2005}.  

In this Letter we consider the simple experiment
that allows us to prove the fractional continuous media model \cite{PLA05}
for fractal media. 
This simple experiment prove that the fractional integrals 
can be used to describe fractal media.


The fractal medium cylinder with the mass $M$, radius $R$,
and the fractal mass dimension,  has the 
moment of inertia in the form
\[ I^{(\alpha)}_z=\frac{\alpha}{\alpha+2} MR^2 , \]
where $\alpha$ is a fractal mass dimension of cross-section of
the cylinder ($1<\alpha \le 2$). The parameter $\alpha$ can be calculated 
by box counting method for the cross-section of the cylinder.
As the result, we get that two homogeneous cylinders 
with equal masses and radiuses can have the unequal moments of inertia. 
These cylinders have the different properties as Maxwell pendulums.
The periods of oscillation for these pendulums are connected by the equation
\[ \Bigl({T^{(\alpha)}_0}/{T^{(2)}_0}\Bigr)^2=
\frac{4(\alpha+1)}{3(\alpha+2)} . \]
where $T^{(\alpha)}_0$ is a period of oscillation of 
fractal medium Maxwell pendulum.
For example, the experiment can be realized by using the sandstone.
Note that Katz and Thompson \cite{KT} presented experimental 
evidence indicating that the pore spaces of a set of sandstone samples 
are fractals and are self-similar over three to four orders of magnitude 
in length extending from 10 angstrom to 100 $\mu m$. 
The deviation $T^{(\alpha)}_0$ from $T^{(2)}_0$ is no more that 6 per cent.
Therefore the precision of the experiments must be high.


\end{document}